\documentstyle[sprocl,psfig]{article}

\def\Journal#1#2#3#4{{#1} {\bf #2}, #3 (#4)}

\def\NIMA{{\em Nucl.\,Instrum.\,Methods}\,A}

\def\NPA{{\em Nucl\, Phys.}\,A}
\def\PLB{{\em Phys.\,Lett.}\,B}
\def\PRL{\em Phys.\,Rev.\,Lett.}
\def\PRD{{\em Phys.\,Rev.}\,D}
\def\ZPC{{\em Z.\,Phys.}\,C}

\def\be{\begin{equation}}
\def\ee{\end{equation}}
\def\bea{\begin{eqnarray}}
\def\eea{\end{eqnarray}}

\begin{document}

\title{${\bf B_{c}} $ MESONS AS A SIGNAL OF DECONFINEMENT}

\author{LEWIS P. FULCHER}

\address{Department of Physics and Astronomy, Bowling Green State
University \\ Bowling Green, OH 43403 \\ email: fulcher@newton.bgsu.edu}

\author{JOHANN RAFELSKI, ROBERT L. THEWS}

\address{Department of Physics, University of Arizona, Tucson, AZ 85721\\
email: rafelski@physics.arizona.edu, thews@physics.arizona.edu }

\maketitle\abstracts{We investigate the fate of bottom quarks 
produced in
heavy-ion collisions at RHIC. Examining both the direct capture
of a charmed quark, and multi-step processes, 
where the $B_{c} $ meson is formed
in a sequence of quark capture and exchange reactions, we find:
a) that a sufficiently
high number of $B_{c}$'s will be produced to generate 
a detectable tri-lepton signal, and b) that the production rate 
of  $B_{c}$'s is highly sensitive to the properties
of the deconfined source. A flavor-independent potential
model, which includes color screening effects, is used to study the
propagation of a $B_{c}$ in a quark-gluon fireball and to
compare this behavior with that of $J/\psi $ 
mesons.}

\begin{center}
\vskip -12cm
{\bf Presented at the APS Centennial meeting 
ATLANTA, March 1999\\ Heavy Ion Minisymposium, to appear in
proceedings,\\ R. Seto, ed., World Scientific, (Singapore 1999).}
\vskip 10.cm
\end{center}

\section{Introduction}
We argue here that RHIC energies should create physical conditions
conducive to the use of $B_{c} $ mesons as a new observable of deconfinement
and quark-gluon plasma (QGP) formation, in spite, and also because
of the fact that elementary
processes lead to the small branching ratio \cite{KR98},
\begin{equation}\label{RBc}
R_{Bc}^{\rm elem}=\frac{B_c+B_c^*}{b\bar b}\simeq 10^{-4}\mbox{--}10^{-5}\,.
\end{equation}
The fast electromagnetic decay of  the vector meson $B_c^*$ allows us 
to  ignore here the distinction between the scalar and vector mesons.
The CDF collaboration  recently announced the
discovery of the $B_{c} $ system~\cite{abe198,abe298}. 
Their announcement was based on
careful analysis of 20 events with characteristic three-charged particle
tracks expected from the decay  $B_{c}^{\pm } \rightarrow J/\psi + 
l^{\pm} + X. $
The CDF collaboration determined values for both the mass
and the lifetime of the ground state,
\begin{equation}\nonumber
M(B_{c}) = 6400 \pm 390 \pm 130 \; {\rm MeV}, \hspace{0.5in} 
\tau(B_{c}) = 0.46^{+0.18}_{-0.16} \pm 0.03 \; {\rm ps}.
\end{equation}

Allowing that either the c quark or the $\bar{\rm b}$ quark can decay
while the other is simply a spectator, or that the c quark and the $\bar{\rm b }$
can annihilate into an intermediate vector boson, the weak decay width of
the ground state of the $B_{c} $ is the sum of three terms, namely,
\begin{equation}
\Gamma (B_{c} \rightarrow X) = \Gamma (\bar{b} \rightarrow X) + \Gamma (
	c\rightarrow X ) + \Gamma(\mbox{annih}).
\end{equation}
Thus our potential model calculation of section \ref{bcbound} 
yields $\tau(B_{c}) = 0.36 \pm 0.05 \; ps$,
which agrees with the CDF result. 
Our calculation of the mass of $B_{c} $ described below 
in section \ref{bcbound} is also in good
agreement with the CDF measurement.                                               

This lifetime allows $B_{c} $ mesons with sufficient
transverse velocity to escape the beam centroid and thus produce a
secondary tri-lepton vertex, which serves as a distinctive feature for
identifying the decay. It is noteworthy that the lifetimes of all
$B_{c} $ states below threshold are much longer than their counterparts
in charmonium or the upsilon system because flavor conservation prevents
decays through gluon annihilation diagrams. 

In a QGP environment at RHIC
the formation of this `stable' exotic 
$B_{c}$ meson is facilitated 
by the ready availability of charmed quarks. We explore here 
the production enhancement of $B_c$ with a view at both a study 
of its properties and as diagnostic tool of the QGP state.

\section{${\bf B_{c}} $ Production and Survival in a Nuclear Fireball}
Although the energy available to the individual nucleons in a typical
RHIC experiment is not as high as that at Fermilab, the formation of
QGP presents an environment that should enhance the
production of bound $B_{c} $ states, since both b and c quarks should
be able to propagate freely. At typical RHIC energies parton fusions,
\begin{equation}
g + g \rightarrow Q + \bar{Q}, \hspace{0.7in} q + \bar{q} \rightarrow
	Q + \bar{Q},
\end{equation}
should produce
a supply of propagating charmed and occasional b and $\bar{\rm b} $ quarks 
with energies less
than 10 GeV. The recent calculation of Mustafa {\em et al.\/}\cite{mu98}
shows that gluon bremsstrahlung provides a very efficient mechanism for
energy loss and that most of the heavy quarks with energies in this range
should stop in a typical distance of 2 fm. An adequate flux of charmed
quarks would lead  to the production of $B_{c} $ states by an 
inelastic scattering accompanied by  gluon bremsstrahlung. 

Since the strange quark
flux is probably about 100 times larger than the charmed quark flux,
$B_{s} $ states should be formed much more frequently than $B_{c} $ states.
Because of the tight binding of the $B_{c} $ states,
quark exchange $B_s+c\leftrightarrow B_c+s$ 
is an exothermic reaction, and we expect this two-step
process to be also an important mechanism for production of $B_{c} $'s.

Using a total
b$\bar{\rm b} $ production cross section of 1.7 microbarns and an overlap
function appropriate for a head-on Au-Au collision at 100 A GeV,
we get a b$\bar{\rm b} $ production rate of 0.05 pairs per central collision.
The abundance of $c\bar c$-pairs has been estimated to be 200 times
bigger, and our population evolution calculations yield a much greater fraction
$R_{Bc}$ than has been obtained for the one step reactions, Eq.\,(\ref{RBc}). 
A further enhancement of the
relative $B_c$-yield arises from multi-step processes described above,
operating in particular 
in the final stages of the QGP evolution. Our calculations show in 
 presence of plasma a major change in the value $R_{Bc}$, Eq.\,(\ref{RBc}),
and thus a major increase in the absolute yield of the $B_c$ mesons.

\begin{center}
\begin{figure}[htb]
\vskip -0.2cm
\hskip 0.5cm\psfig{figure=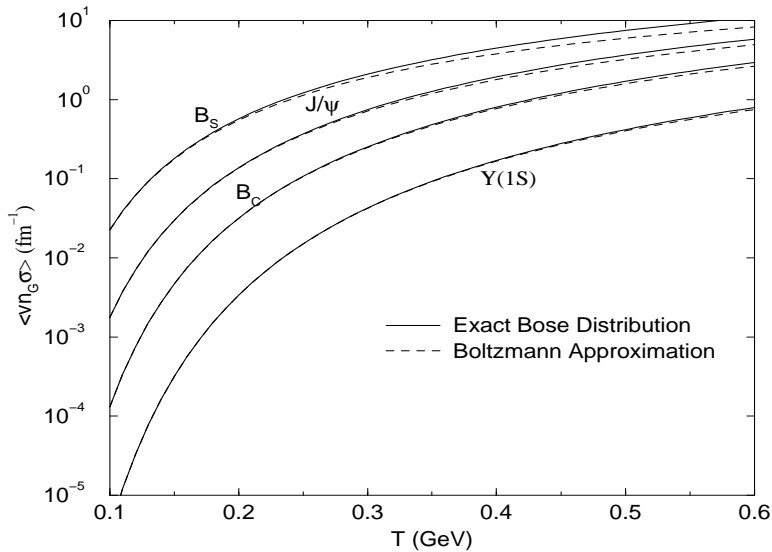,height=2.9in,width=4.0in}
\vskip -0.5cm
\caption{Quarkonium dissociation rates as functions of temperature.
\label{F1}}
\vskip -0.5cm
\end{figure}
\end{center}

We cannot describe here all the intricacies of the population evolution
equations that establish the final production rates of the $B_c$\,. We 
mentioned already that there are several channels of production. Beyond that 
the issue is to assess the prospects for the survival of the
$B_{c} $ state in the hot parton environment. Using Kharzeev's 
approach \cite{kh98}, we have calculated the quarkonium dissociation rates
for $J/\psi $, $B_{c} $ and $\Upsilon(1S) $, as shown in Fig.\,\ref{F1}. 
The curves
there indicate that $B_{c} $ should travel 5 to 10 times further than
$J/\psi $ at $T = 300$\,MeV.
It is worth noting that the probability of dissociation
of the $B_{c} $ at $T=400$\,MeV is about the same as that of 
$J/\psi $ at $T=300$\,MeV. 

\section{Potential Model Approaches to the ${\bf B_{c}} $ System}
\label{bcbound}    
Another way to understand the unusual stability of the $B_c$ mesons
in the plasma phase is to consider its bound state structure.
The rich spectra of bound states of charmonium and the upsilon system
below the flavor thresholds~\cite{pdg98} provide enough information
to determine the parameters of a nonrelativistic potential model for
heavy quarks. The assumption of flavor independence extends this model
to the $B_{c} $ system. 
We describe here one such recent calculation~\cite{fu98} 
incorporating a model of a running coupling constant effects in the
central potential and the full radiative one-loop expressions supplemented
by the Gromes consistency condition to incorporate non-perturbative
effects in the spin-dependent potentials: these 
calculations provided an excellent fit of the upsilon levels 
(avg. dev. = 4.3\,MeV), a good fit of the charmonium levels
(avg. dev. = 19.9\,MeV) and a good account of the leptonic widths below
threshold. The model predicts a ground state energy of
$M(B_{c}) = 6286 ^{+15}_{-6}$\,MeV and that this state is 820\,MeV below
the threshold for heavy-light flavored meson production. Notably, the ground
state of $B_{c} $ is expected to be about 150\,MeV more tightly bound than
the ground state of charmonium, and one thus would expect it to be more likely to
survive QGP environments with temperatures around 300\,MeV.

\begin{center}
\begin{figure}[ht]
\vskip 0.7cm
\hskip 0.7cm\psfig{figure=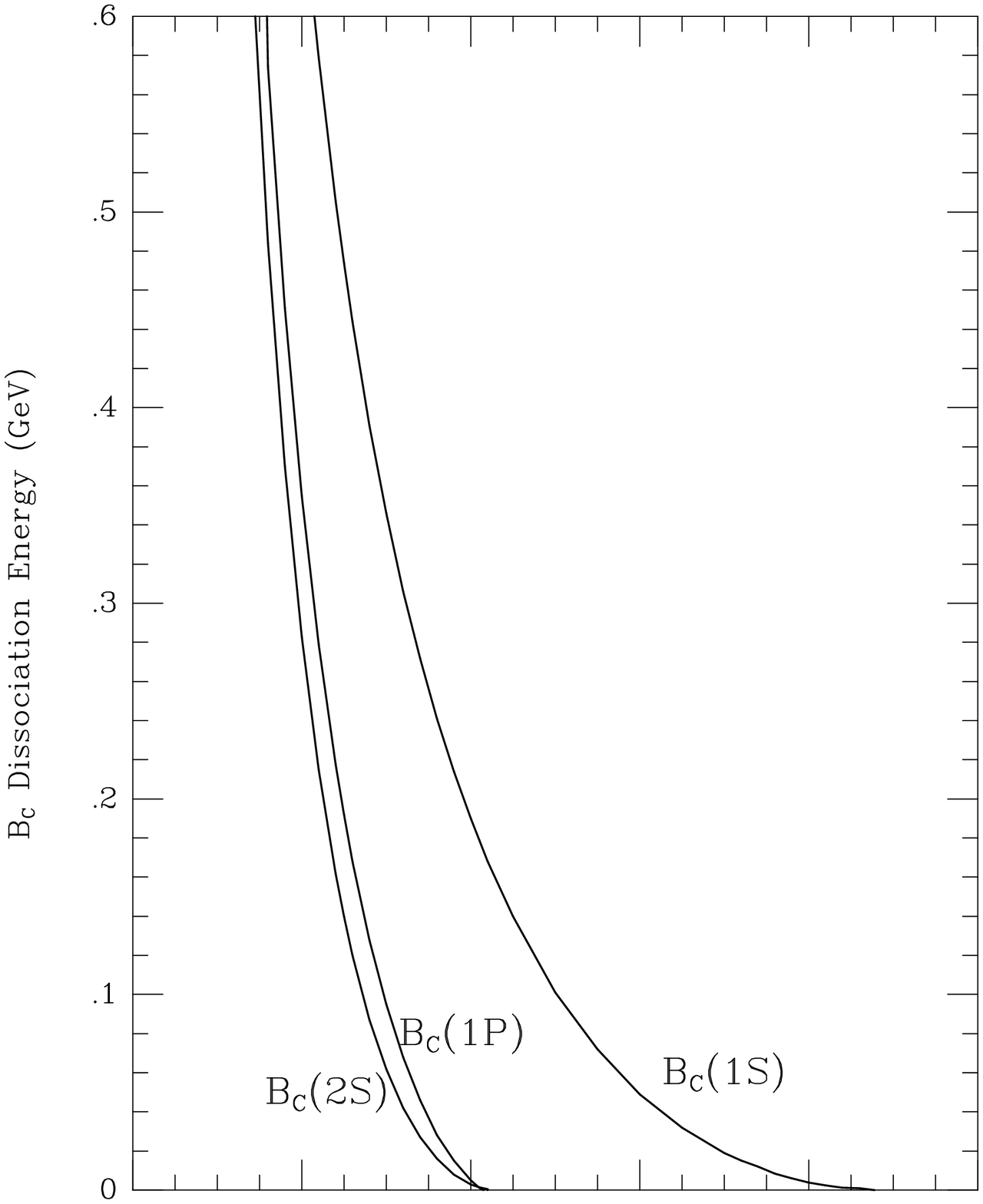,height=2.0in,width=4.0in}
\vskip -0.25cm
\hskip 0.7cm\psfig{figure=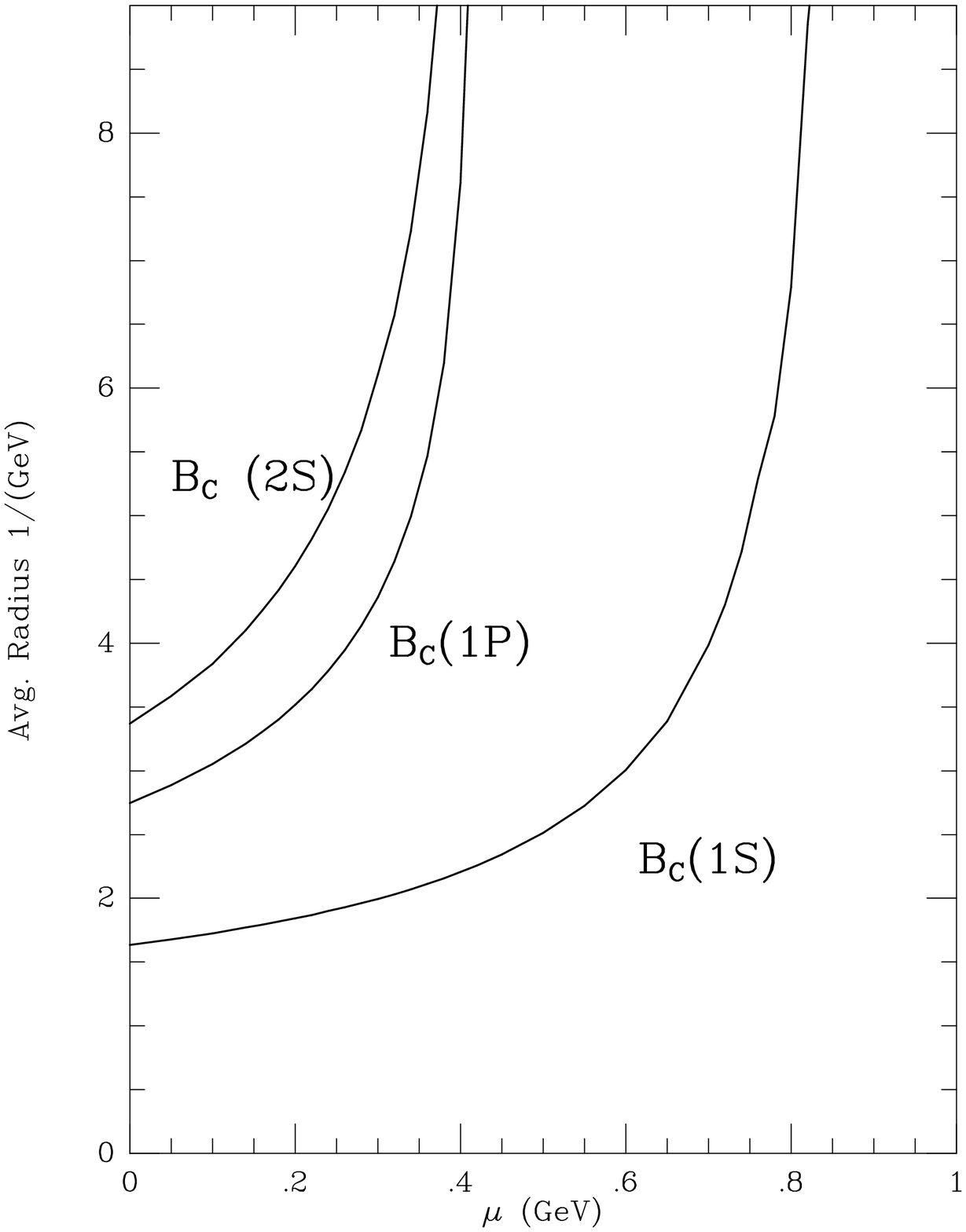,height=2.0in,width=4.0in}
\vskip -0.2cm
\caption{Top: $B_{c} $ dissociation energies, bottom: $B_{c} $ radii 
      as functions of inverse screening length.\label{F2}}
\vskip -1cm
\end{figure}
\end{center}

To extend the potential model concept to QGP, one needs
a means to consider the effects of color screening. Thus, we follow the
the approach of Karsch, Mehr and Satz \cite{ka88}, who introduce color
screening effects with an exponential damping factor. Their analysis
begins with a Coulomb plus linear potential, which describes the $T = 0 $
limit. Such a potential is known to give an adequate account of heavy-quark
spin-averaged energies \cite{fu94}. Their inverse Debye screening length
parameter $\mu(T) $ describes the temperature dependence of color screening. 
It also allows for deconfinement by parameterizing the melting of the string
tension. The Karsch-Mehr-Satz form for the central potential is 
\begin{equation}\label{VKMS}
V_{\rm KMS}(\mu(T), R) = A \frac{\left( 1 - e^{-\mu r} \right) }{\mu } - \kappa
	\frac{ e^{-\mu r} }{r}.
\end{equation}
   From the large distance limit, the energy required to
dissociate the system is
\begin{equation}
E_{dis}(\mu ) = m_{b} + m_{c} + \frac{A}{\mu } - E_{bc}(\mu )\,,
\end{equation}
where the  binding energy $E_{bc}(\mu )$ is obtained solving 
the  Schr\"odinger equation with $V_{\rm KMS}(\mu(T), R)$, Eq.\,(\ref{VKMS}).
Deconfinement is achieved when the dissociation energy vanishes, thus defining
the critical value of $\mu $. Our results for the dissociation
energies of the three lowest $B_{c}$ states are shown in the top portion of
 Fig. \ref{F2}. There one
sees that $\mu_{crit}(1S) = 840$\,MeV, almost 200\,MeV higher than
the corresponding quantity for $J/\psi $. The $\mu $ dependence of the sizes
of these 3 lowest $B_{c} $ states are shown in the bottom portion of
 Fig. \ref{F2}.  For any given value
of $\mu $, the dissociation energy for the $B_{c} $ state is larger
than the value for the corresponding state in charmonium, and the average radius
is smaller.

\section{Outlook and Conclusions}    
The computation of the ratio $R_{Bc}$, Eq.\,(\ref{RBc}), in the deconfined
QGP environment poses a challenge, and our estimates yield at present a few 
percent. In any case a very 
significant, by a factor 100--1000, and observable 
enhancement over elementary processes
is to be expected. Based on the 
known parameters of the RHIC we have worked out some interesting numbers
shown in table~\ref{yields}. The reason that the chemical non-equilibrium
for charm is leading to a greater yield is the fact that the directly produced
charm is here dominating the thermal production. There is practically no
re-annihilation of charmed quarks as plasma cools down and hence the significant
increase of the tri-lepton channel decay yield of the $B_c$. 
\begin{table}[ht]
\caption{\label{yields} RHIC yields for several heavy  quark systems.
}\begin{center} 
\begin{tabular}{|l|cc|} \hline\hline
observable&first year&design luminosity\\
\hline
$c\bar c$-pairs & $2.8\,10^8$ & $6.5\,10^9$ \\
$b\bar b$-pairs & $1.2\,10^6$ & $3.2\,10^7$ \\
$J/\Psi\to\mu^+\mu^-$ & $1.6\, 10^5$ & $3.9\,10^6$ \\
$\Upsilon(1s)\to\mu^+\mu^-$ &140 & 3800 \\
\hline
$B_c\to J/\Psi + l\nu $& \hspace*{2.4cm}no & QGP\hspace*{2.3cm} \\
\hspace*{.8cm} $\to \mu^+\mu^-+ l\nu $ & 0.05--0.18 & 1.5--4.9 \\
\hline
 $B_c\to J/\Psi + l\nu $& therm+chem. equil. & QGP\hspace*{2.4cm}\\
\hspace*{.8cm} $\to \mu^+\mu^-+ l\nu $ & 18 & 490 \\
\hline
 $B_c\to J/\Psi + l\nu $& \hspace*{.8cm}only prim. $c\bar c$ & QGP$(T=400$\,MeV)\\
\hspace*{.8cm} $\to \mu^+\mu^-+ l\nu $ & 235 & 6400 \\
\hline
 $B_c\to J/\Psi + l\nu $& \hspace*{.8cm}only prim. $c\bar c$ & QGP$(T=200$\,MeV)\\
\hspace*{.8cm} $\to \mu^+\mu^-+ l\nu $ & 1000 & 27,000 \\
\hline
\hline
\end{tabular} 
\end{center} 
\vspace{-0.4cm}
\end{table} 

Our conclusions are:\\
(1) The larger binding energy of the $B_{c} $ system and the 
    availability of the primary-interaction charmed quarks facilitates 
    production and stability of the $B_c$ meson;\\
(2) The  $B_{c} $  should be observable at RHIC  (as well as at LHC-Alice),
     and the study  of the elementary properties of the $B_c$ mesons 
    appears possible;\\
(3)  The dynamics of  the $B_{c} $ system in plasma (given its smaller
     size and greater binding) is different from that of $J/\psi $, and 
     thus $B_{c} $ should produce information that is complementary to 
      that produced by $J/\psi $, and $\Upsilon(1s)$;\\
(4)   There is potential for a `smoking gun' evidence of deconfinement 
      indicated by a significant enhancement of 
      $R_{Bc}$, and thus  it  is important to evaluate this quantity 
     theoretically and measure it experimentally in  QGP 
      environment.\vskip0.4cm

\noindent {\large {\bf References}}

\end{document}